\documentclass[reqno,centertags,12pt]{amsart}
\usepackage{amsmath,amsthm,amscd,amssymb}
\usepackage{latexsym}

\usepackage {verbatim}

\newcommand{\N}{{\mathbb{N}}}

\newcommand{\R}{{\mathbb{R}}}

\newcommand{\C}{{\mathbb{C}}}

\newcommand{\Hmm}[1]{\leavevmode{\marginpar{\tiny%
$\hbox to 0mm{\hspace*{-0.5mm}$\leftarrow$\hss}%
\vcenter{\vrule depth 0.1mm height 0.1mm width \the\marginparwidth}%
\hbox to 0mm{\hss$\rightarrow$\hspace*{-0.5mm}}$\\\relax\raggedright #1}}}

\newcommand{\ol}{\overline}

\newcommand{\wti}{\widetilde  }

\newcommand{\ran}{\text{\rm{ran}}}

\newcommand{\ess}{\text{\rm{ess}}}

\newcommand{\hatt}{\widehat}
\newcommand{\beq}{\begin{equation}}
\newcommand{\eeq}{\end{equation}}
\newcommand{\bdm}{\begin{displaymath}}
\newcommand{\edm}{\end{displaymath}}
\newcommand{\ba}{\begin{align}}
\newcommand{\ea}{\end{align}}
\newcommand{\veps}{\varepsilon}

\newcommand{\supp}{\mathrm{supp}}               
\newcommand{\disc}{\mathrm{disc}}

\newcommand{\vare}{\varepsilon}

\newcommand{\re}{\mathrm{Re}}
\newcommand{\im}{\mathrm{Im}}

\newcommand{\norm}[1]{\Vert#1\Vert}  

\newcommand{\nul}{\operatorname{nul}}
\newcommand{\de}{\operatorname{def}}
\newcommand{\ind}{\mathrm{ind}}

\DeclareMathOperator{\Real}{Re}

\allowdisplaybreaks

\newcommand{\calC}{\mathcal{C}}
\newcommand{\calD}{\mathcal{D}}

\newcommand{\calH}{\mathcal{H}}

\newcommand{\calN}{\mathcal{N}}

\newtheorem{thm}{Theorem}

\newtheorem{lemma}[thm]{Lemma}
\newtheorem{corollary}[thm]{Corollary}

\theoremstyle{definition}

\newtheorem{remark}[thm]{Remark}

\newcounter{theoremi}[thm]

\newcommand{\itemthm}{\refstepcounter{theoremi} {\rm(\roman{theoremi})}{~}}

\numberwithin{thm}{section}
\numberwithin{equation}{section}

\def\defeq{{\mathchoice{%
\mathrel{\mskip\thickmuskip\raise.35pt\hbox{$\mathord{\displaystyle:}$}%
\hbox{$\mathord{\displaystyle=}$}\mskip\thickmuskip}}{%
\mathrel{\mskip\thickmuskip\raise.35pt\hbox{$\mathord{\displaystyle:}$}%
\hbox{$\mathord{\displaystyle=}$}\mskip\thickmuskip}}{%
\mathrel{\mskip.25\thinmuskip\raise.25pt\hbox{$\mathord{\scriptstyle:}$}%
\hbox{$\mathord{\scriptstyle=}$}\mskip.25\thinmuskip}}{%
\mathrel{\mskip.1\thinmuskip\raise.1pt\hbox{$\mathord{\scriptscriptstyle:}$}%
\hbox{$\mathord{\scriptscriptstyle=}$}\mskip.1\thinmuskip}%
}}}
\def\eqdef{{\mathchoice{%
\mathrel{\mskip\thickmuskip\hbox{$\mathord{\displaystyle=}$}%
\raise.35pt\hbox{$\mathord{\displaystyle:}$}\mskip\thickmuskip}}{%
\mathrel{\mskip\thickmuskip\hbox{$\mathord{\displaystyle=}$}%
\raise.35pt\hbox{$\mathord{\displaystyle:}$}\mskip\thickmuskip}}{%
\mathrel{\mskip.25\thinmuskip\hbox{$\mathord{\scriptstyle=}$}%
\raise.25pt\hbox{$\mathord{\scriptstyle:}$}\mskip.25\thinmuskip}}{%
\mathrel{\mskip.1\thinmuskip\hbox{$\mathord{\scriptscriptstyle=}$}%
\raise.1pt\hbox{$\mathord{\scriptscriptstyle:}$}\mskip.1\thinmuskip}%
}}}
\def\halmos{{%
\hspace*{\fill}\hbox to 18pt {\hfill\vrule width 9pt height 9pt depth
0pt}%
}\par\ignorespaces}

\begin{document}

\title[Exponential Decay]{Exponential decay of eigenfunctions and generalized
eigenfunctions of a non self-adjoint matrix Schr\"odinger operator
related to NLS}
\author[D.~Hundertmark and Y.-R.~Lee]{Dirk Hundertmark and Young-Ran~Lee}
\address{School of Mathematics, Watson Building, University of Birmingham, Edgbaston, Birmingham, B15 2TT, UK.
    On leave from Department of Mathematics, Altgeld Hall, University of Illinois at Urbana-Champaign,
    1409 W.~Green Street, Urbana, IL 61801.}%
\email{hundertd@maths.bham.ac.uk and dirk@math.uiuc.edu}%
\address{Department of Mathematics, Altgeld Hall, University of Illinois at Urbana-Champaign,
                1409 W.~Green Street, Urbana, IL 61801.}%
\email{yrlee4@math.uiuc.edu}

\thanks{D.H.~supported in part by NSF grant DMS--0400940}
\thanks{\copyright 2007 by the authors. Faithful reproduction of this article,
        in its entirety, by any means is permitted for non-commercial purposes}
\keywords{Non-linear Schr\"odinger equation, decay of eigenfunctions}
\subjclass[2000]{35B20, 35B40, 35P30 }
\date{22 March 2007, revised version exponential13.tex }

\begin{abstract}
We study the decay of eigenfunctions of the non self-adjoint matrix operator
$\calH = \left(
\begin{smallmatrix}
 -\Delta +\mu+U & W \\-W & \Delta -\mu -U
\end{smallmatrix}
\right)$, for $\mu>0$, corresponding to eigenvalues in the strip $-\mu<\re E <\mu$.
\end{abstract}

\maketitle

\section{Introduction}\label{sec1}
For some positive $\mu$, we consider the system
    \begin{equation}\label{eq:calH}
    \calH :=\left(  \begin{array}{cc}
      -\Delta+\mu+U & W \\
      -W & \Delta-\mu-U
      \end{array}
    \right),
    \end{equation}
with real-valued functions $U$ and $W$. We will impose some weak
conditions on $U$ and $W$ which insure that $\calH$ is a closed
operator on the domain $\calD(\calH)= H^2(\R^d,\C^2)$. The
unperturbed operator $\calH_0$, where $U=W=0$, is given by
 \bdm
  \calH_0 :=\left(  \begin{array}{cc}
      -\Delta+\mu & 0 \\
      0 & \Delta-\mu
      \end{array}
    \right).
 \edm
Note that $\calH_0$ is a self-adjoint operator on the domain
$H^2(\R^d,\C^2)$ and, by inspection, the spectrum of $\calH_0$
equals $\sigma(\calH_0)=(-\infty,-\mu]\cup [\mu,\infty)$.

 Our assumptions on $U$ and $W$ are
\begin{itemize}
\item[A.] $U$ and $W$ are $-\Delta$-bounded with relative bound zero. That is,
    the domains $\calD(U)$ and $\calD(W)$, as multiplication operators with
    real-valued functions, contain
    the Sobolev space $H^2(\R^d) = \calD(-\Delta)$ and for all $\vare>0$ there
    exists a $C_\vare$ such that
\bdm
    \norm{X g}_2\le \vare\norm{-\Delta g}_2 +C_\vare\norm{g}_2 \quad \text{for } X=U,W.
\edm \item[B.] $U$ and $W$ decay to zero at infinity.
\end{itemize}

Let us explain how such a non-symmetric system naturally arises in
the stability/instability study of solutions of the non-linear
Schr\"odinger equation (NLS): The NLS is a non-linear evolution
equation of the form
 \beq\label{eq:NLS}
 i\partial_t \psi =
 -\Delta \psi - F(|\psi|^2)\psi \quad \text{ on } \R^d
 \eeq
for some real-valued non-negative function $F$, for example, $F(s)=
s^\sigma$  with $\sigma>0$. Making the ansatz $\psi(t,x) =
e^{it\mu}\phi(x)$, with $\mu>0$, one gets the time independent NLS
 \beq\label{eq:NSLtime-independent}
  (-\Delta +\mu -F(|\phi|^2))\phi = 0 .
 \eeq
Now let $\phi$ be a solution of \eqref{eq:NSLtime-independent}. If
$\phi>0$ one often calls it the non-linear ground state, but we will
not make this requirement. Perturbing $\psi$ a bit, one makes the
ansatz $\psi= e^{it\mu}(\phi+R)$ and gets the equation
 \bdm
 i\partial_t R
 =
 \big[-\Delta +\mu -(F(|\phi|^2) + F'(|\phi|^2)|\phi|^2) \big] R
  -F'(|\phi|^2) \phi^2 \ol{R} + N
 \edm
where $N$ is a term quadratic in $R$.   Note that $\phi$ can always
be chosen to be real-valued, however, we do not need to make this
assumption. The above can be written as the system,
 \beq\label{eq:system}
 i\partial_t \left(\begin{array}{r} R\\ \ol{R} \end{array}\right)
 =
 \calH \left(\begin{array}{r} R\\ \ol{R} \end{array}\right) +\calN
 \eeq
where $\calN$ is a term quadratic in $R$, $\calH$ is as in
\eqref{eq:calH}, and the potentials are given by $U= - F(|\phi|^2)-
F'(|\phi|^2)|\phi|^2$ and $W= - F'(|\phi|^2)\phi^2$. Hence the
non-linear system \eqref{eq:system} describes the time behavior of a
perturbation $R$ of the NLS around a stationary solution $\phi$.

  Since $\calN$ is quadratic in $R$, one sees that to first
order in the perturbation $R$, the spectral properties of systems
like the one in \eqref{eq:calH} determine the linear
stability/instability properties of \eqref{eq:system} and hence the
linear stability/instability of stationary solutions of the NLS
\eqref{eq:NLS}. For this reason, the study of the spectral
properties of operators given by \eqref{eq:calH} has received
renewed interest in recent years, see, for example,
\cite{BP,CGNT06,CPV,ES,RSS,Schlag,VP}.

\begin{remark}
\itemthm \label{rem:resolvent-bound} Assumption A is equivalent to
 \bdm
 \lim_{\lambda\to\infty} \norm{X(-\Delta +i\lambda)^{-1}} = 0
 \edm
for $X=U,W$, see, for example, \cite{CFKS,HislopSigal,ReSi1}. Note
that because of assumption A, the operator $\calH$ is a closed
operator on its domain $\calD(\calH)= \calD(\calH_0)=
H^2(\R^d,\C^2)$.\\[0.2em]
\itemthm Assumption A is fulfilled, if $U$ and $W$ obey certain
local uniform $L^p$-conditions, $U,W\in
L^p_{\mathrm{loc,unif}}(\R^d)$ with $p=2$ for $d\le 3$ and $p>d/2$
if $d\ge 4$, or, slightly more generally, if they are in the
Stummel class $S_d$, see \cite{CFKS,Schechter,Stummel}. \\[0.2em]
\itemthm We want to stress the fact that we do not make any assumptions on how
fast $U$ and $W$ decay, only that they tend to zero at infinity.
\end{remark}

Our first result deals with the essential spectrum of systems like
\eqref{eq:calH}. Since there are several non-equivalent definitions
for the essential spectrum of non-selfadjoint operators, let us
discuss these a little bit in more detail: Let $T$ be an arbitrary
closed operator on a Hilbert space. Its resolvent set $\rho(T)$
consists of all $z\in\C$ such that $T-z$ is boundedly invertible.
Its spectrum is given by $\sigma(T)= \C\setminus\rho(T)$.  A closed
operator $T$ is Fredholm, if its range is closed and both the kernel
and co-kernel, the orthogonal complement of its range, are
finite-dimensional. Its index is the difference of the dimensions of
its kernel and co-kernel, $\text{ind}(T)= \text{nul}(T)-\text{def}$
$T$, where $\text{nul}(T)=\dim\ker(T)$ and $\text{def}(T)= \dim
\ran(T)^\bot$. $T$ is semi-Fredholm, if its range is closed and
either its kernel or co-kernel is finite-dimensional. Consider the
following sets
 \begin{itemize}
 \item $\Delta_1(T)= \{z \in \C\vert\, T-z \text{ is semi-Fredholm} \}$.
 \item $\Delta_2(T)= \{z\in \C\vert\, T-z \text{ is semi-Fredholm and
 }\text{nul}(T-z)<\infty \}$.
 \item $\Delta_3(T)= \{z \in \C\vert\, T-z \text{ is Fredholm} \}$.
 \item $\Delta_4(T)= \{z \in \C\vert\, T-z \text{ is Fredholm with } \text{ind}(T-z)=0
 \}$.
 \item $\Delta_5(T)= \{z \in \Delta_4(T)\vert  \text{ a deleted neighborhood of $z$ is in }\rho(T)\}$.
 \end{itemize}
Note that $\rho(T)=\{z\in\C\vert\, T-z \text{ is Fredholm with
}\text{nul}(T-z)= \text{def}(T-z)=0\}$. Thus all sets defined above
contain the resolvent set $\rho(T)$ and possible definitions for the
essential spectrum,  in terms of Fredholm properties, are given by
 \bdm
 \sigma_{\ess,j}(T)= \C\setminus \Delta_j(T), \quad j=1,\ldots,5 .
 \edm

\begin{remark}
\itemthm These definitions are taken from page 40 in
\cite{EdmundsEvans}, see also \cite{GustafsonWeidmann}. The first
one is the one used by Kato, see page 243 in \cite{Kato}, the fifth
was introduced by Browder, \cite{Browder}, see also the discussion
in Appendix \ref{app:B}.  \\[0.2em]
\itemthm \label{rem:alternativesigmaess5} Theorem IX-1.5 in
\cite{EdmundsEvans} shows that
 $\sigma_{\text{ess},5}(T)$ is the union of
 $\sigma_{\text{ess},1}(T)$ with all components of $\C\setminus
 \sigma_{\text{ess},1}(T)$ which do not intersect the resolvent
 set. \\[0.2em]
\itemthm For self-adjoint operators, all definitions above coincide. In
general, one has the inclusions
 \bdm
  \sigma_{\ess,1}(\calH) \subset \sigma_{\ess,2}(\calH)
  \subset \sigma_{\ess,3}(\calH) \subset\sigma_{\ess,4}(\calH)
  \subset \sigma_{\ess,5}(\calH) \subset \sigma(\calH)
 \edm
since $\Delta_j(T)$ is a decreasing sequence of sets containing the
resolvent set $\rho(T)$. All of the above inclusion can be strict,
see the discussion in \cite{GustafsonWeidmann}. \\[0.2em]
\itemthm \label{rem:essentialspectrum6} Another natural definition,
in the spirit of the essential spectrum for self-adjoint operators,
is to define the essential spectrum as the complement (in the
spectrum) of the discrete spectrum. More precisely, if we denote by
$\sigma_{\text{disc}}(T)$ the set of all isolated points
$\lambda\in\sigma(T)$ with finite \emph{algebraic} multiplicity.
Then the essential spectrum should be given by
$\sigma(T)\setminus\sigma_{\text{disc}}(T)$. This definition of
essential spectrum is introduced on page 106 in \cite{ReSi4}. In
fact, it coincides with the fifth one,
 \beq\label{eq:equivalence}
 \sigma_{\text{ess},5}(T) =
 \sigma(T)\setminus\sigma_{\text{disc}}(T).
 \eeq
We could not find any proof of this in the literature and, for the
convenience of the reader, give a proof of this in
Appendix \ref{app:B}.\\[0.2em]
\itemthm It might be surprising that in our case all of the above
five definitions of essential spectrum coincide, as the following
theorem shows.
\end{remark}

\begin{thm}\label{thm:spectrum} Under the above conditions on $U$
and $W$, one has
 \bdm
 \sigma_{\ess,j}(\calH) = \sigma(\calH_0)=
 (-\infty,-\mu]\cup [\mu,\infty)
 \qquad\text{ for } j=1,2,3,4,5,
 \edm
and the spectrum of $\calH$ outside of its essential spectrum
consists of a discrete set of eigenvalues of finite algebraic
multiplicity.

Moreover, $\sigma(\calH)$ is symmetric under reflection along the real and
imaginary axes, that is, $\sigma(\calH)= -\sigma(\calH)$ and
$\sigma(\calH^*)= \sigma(\calH)$.
\end{thm}
\begin{remark}
\itemthm The only condition needed on $V= \left
(\begin{smallmatrix} U& W \\ -W &-U\end{smallmatrix}\right)$ is that
$V$ is relatively $\calH_0$-compact, which is the case if $U$ and $W$
are relatively Laplacian compact. \\[0.2em]
\itemthm Because of Theorem \ref{thm:spectrum}, there is no need to
distinguish between the different definitions for the essential
spectrum in the following. \\[0.2em]
\itemthm \label{rem:gen-eigenspace}
Since $\calH$ is not self-adjoint, it can  happen that, for some
eigenvalue $z$, $\ker((\calH-z)^2)\not = \ker(\calH-z)$, that is,
$\calH$ can possess generalized eigenspaces (a non-trivial Jordan
normal form). However, the generalized eigenspace stabilizes, that
is, for any $z\in\sigma(\calH)\setminus\sigma_{\ess}(\calH)$ there
is a $k\in\N$ with $\ker(\calH-z)^{m+1}= \ker(\calH-z)^m$ for all
$m\ge k$, see the proof of Theorem \ref{thm:spectrum}. In the
application to the non-linear Schr\"odinger equation, this typically
happens at $z=0$, see \cite{Weinstein1, Weinstein2, Schlag, RSS}.\\[0.2em]
\itemthm Under the so-called positivity condition,
 \bdm
 L_-\defeq -\Delta +\mu +U -W \ge 0,
 \edm
one has $\sigma(\calH)\subset \R\cup i\R$ and each eigenvalue with
$z\not =0$ has trivial Jordan form,
$\ker((\calH-z)^2)=\ker(\calH-z)$. That is, the generalized
eigenspace for non-zero eigenvalues coincides with the eigenspace.
Under the positivity condition, only $z=0$ can possess a generalized
eigenspace. This is, for example, shown in \cite{BP,RSS} and the
proof carries over to our assumptions on $U$ and $W$.
\end{remark}

Our main goal in this paper is to prove that the generalized eigenfunctions of
the above system with energies in the gap of the essential spectrum decay
exponentially. This is the content of the next theorem.

\begin{thm}\label{thm:Ltwo-exp-decay}
Let $E$ be an eigenvalue of $\calH$ with $-\mu<\Real(E)<\mu$. Then
under the above assumptions on $U$ and $W$, every eigenfunction
and generalized eigenfunction corresponding to $E$ decays
exponentially. More precisely, if $(\calH-E)^k\varphi=0$ for some
$k\in\N$,  we have the $L^2$-decay estimate
\beq\label{eq:Ltwo-decay} e^{(\sqrt{\mu-|\re E|-2\delta})|x|}
\varphi\in L^2(\R^d,\C^2) \qquad \text{for all positive }\delta
<\frac{1}{2}(\mu-|\re E|). \eeq
\end{thm}

\begin{remark} This result improves the exponential decay estimates of
\cite{RSS} in two directions.  First, the authors of \cite{RSS} need
much stronger condition on the off-diagonal part $W$, namely some
exponential decay of $W$. Secondly, they considered only
(generalized) eigenfunctions corresponding to real eigenvalues
within the gap $(-\mu,\mu)$. However, as shown in
\cite{Grillakis89,Grillakis90} and \cite{GSS87}, certain
supercritical non-linearities lead to linearizations of NLS around
the ground state which have a pair of purely imaginary eigenvalues
in addition to their generalized eigenspace at zero, see also Lemma
17 in \cite{Schlag}. Our result shows that no a-priori decay rate
for the matrix potential has to be specified for this. Moreover, the
decay rate is uniform in the imaginary part of the eigenvalues and
explicitly depends only on the positivity of $\mu - |\re E|$.

In the next section we give the proof of Theorem \ref{thm:spectrum}.
Theorem \ref{thm:Ltwo-exp-decay} is proved in Sections
\ref{sec:Ltwo-decay} and \ref{sec:Ltwo-decay-generalized}.
Exponential decay of eigenfunctions is given in Section
\ref{sec:Ltwo-decay} and exponential decay of generalized
eingefunctions in Section  \ref{sec:Ltwo-decay-generalized}.

\vspace{5mm} \noindent {\bf Acknowledgement:} It is a pleasure to
thank Wilhelm Schlag for bringing this type of spectral problem to
our attention. Young-Ran Lee thanks the School of Mathematics at the
University of Birmingham, England, for their warm hospitality.
Furthermore, we would like to thank Des Evans for some discussions
and the unknown referee for numerous comments which, in particular,
helped to improve Theorem \ref{thm:spectrum}.

\end{remark}

\section{Proof of theorem \ref{thm:spectrum}} \label{sec:spectrum}
Using, for example, the Fourier transform, one sees that the
unperturbed operator $\calH_0$ is self-adjoint on $H^2(\R^d,\C^2)$
and that its spectrum is given by
$\sigma(\calH_0)=\sigma_{\ess}(\calH_0)=(-\infty,-\mu]\cup
[\mu,\infty)$. Recall that all different notions of essential
spectrum coincide due to the self-adjointness of $\calH_0$. Since
$U$ and $W$ are Laplacian bounded with relative bound zero and go to
zero at infinity, they are Laplacian compact. In particular, this
implies
for $V= \left (\begin{smallmatrix} U& W \\
-W &-U\end{smallmatrix}\right)$ that
 \bdm
 V(\calH_0-z)^{-1} \quad \text{ is compact}
 \edm
for all $z\in\C\setminus \sigma(\calH_0)=
\C\setminus((-\infty,-\mu]\cup[\mu,\infty))$. Thus, by Theorem
IX-2.1 in \cite{EdmundsEvans},
 \bdm
 \sigma_{\ess,j}(\calH)= \sigma_{\ess}(\calH_0)= (-\infty,-\mu]\cup[\mu,\infty)
 \quad \text{ for } j=1,2,3,4.
 \edm

To prove the claim for $\sigma_{\ess,5}(\calH)$ we need to know a
bit more about the resolvent set of $\calH$. Let us first prove that
the eigenvalues of $\calH$ in $\C\setminus\sigma(\calH_0)$ form a
discrete set with only $\sigma(\calH_0)$ are possible accumulation
points. By Remark \ref{rem:resolvent-bound} and the assumptions on
$V$, we see that
 \beq\label{eq:resolventatinfinity}
 \lim_{\lambda\to\infty} \norm{V(\calH_0+i\lambda)^{-1}} = 0.
 \eeq
Using \eqref{eq:resolventatinfinity}, one sees that $1+
V(\calH_0-z)^{-1}$ is invertible for some complex $z$ and hence the
analytic Fredholm alternative, see \cite{ReSi1}, shows that there
exists a discrete subset $D$ of $\C\setminus\sigma(\calH_0)$ such
that $1+V(\calH_0-z)^{-1}$ is invertible for all $z\in
\C\setminus\sigma(\calH_0)$ which are not in $D$.

This set $D$ is precisely the set of all eigenvalues of $\calH$ in
$\C\setminus\sigma(\calH_0)$. Indeed, by the compactness of
$V(\calH_0-z)^{-1}$, $1+V(\calH_0-z)^{-1}$ is not invertible if and
only if $-1\in\sigma(V(\calH_0-z)^{-1})$. So there exists a
non-trivial $\phi\in L^2(\R^d,\C^2)$ with
 \bdm
 V(\calH_0-z)^{-1}\phi = -\phi .
 \edm
With $\psi = (\calH_0-z)^{-1}\phi$, we can rewrite this as
 \bdm
 (\calH_0 +V)\psi = z\psi ,
 \edm
so $z$ is an eigenvalue of $\calH$ with eigenvector $\psi$. In
addition, reversing the above argument, one sees that if $z\not \in
\C\setminus\sigma(\calH_0)$ is an eigenvalue of $\calH$, then
$-1\in\sigma(V(\calH_0-z)^{-1})$ and hence $z\in D$. So the set $D$
consists of all eigenvalues of $\calH$ in
$\C\setminus\sigma(\calH_0)$.

As a second step, let us show that the resolvent set of $\calH$ is
quite big, it contains $\C\setminus (\sigma(\calH_0)\cup D)$.
Indeed, for any $z\not\in \sigma(\calH_0)\cup D$,
 \begin{multline*}
 (\calH-z) (\calH_0-z)^{-1}\big( 1+V(\calH_0-z)^{-1} \big)^{-1}
 = \\
 (\calH_0-z) (\calH_0-z)^{-1}\big( 1+V(\calH_0-z)^{-1} \big)^{-1}
 + V(\calH_0-z)^{-1}\big( 1+V(\calH_0-z)^{-1} \big)^{-1} = \\
 \big( 1+V(\calH_0-z)^{-1} \big)\big( 1+V(\calH_0-z)^{-1} \big)^{-1}
 = I.
 \end{multline*}
Thus, for those values of $z$, $\calH-z$ is surjective. A similar
calculation shows
 \bdm
 (\calH_0-z)^{-1}\big( 1+V(\calH_0-z)^{-1} \big)^{-1} (\calH-z) = I,
 \edm
so $\calH-z$ is also injective, and hence a bijection if $z\not\in
\sigma(\calH_0)\cup D$. Thus, by the closed graph theorem, $\calH-z$
is boundedly invertible for those values of $z$ with inverse
 \beq\label{eq:resolventformula}
  (\calH-z)^{-1} = (\calH_0-z)^{-1}( 1+V(\calH_0-z)^{-1})^{-1}.
 \eeq
In particular, the resolvent set of $\calH$ contains at least the
set $\C\setminus (\sigma(\calH_0)\cup D )$, where the discrete set
$D$ is the set of eigenvalues of $\calH$ in $\C\setminus
\sigma(\calH_0)$.

Coming back to $\sigma_{\ess,5}(\calH)$, we simply note that, due to
the above, $\C\setminus\sigma_{\ess,1}(\calH)$ is a connected set
which intersects the resolvent set of $\calH$. Hence, by Remark
\ref{rem:alternativesigmaess5},
 \bdm
 \sigma_{\ess,5}(\calH)= \sigma_{\ess,1}(\calH)=
 (-\infty,-\mu]\cup[\mu,\infty)
 \edm
also.

Now we show that the generalized eigenspace corresponding to
eigenvalues $z_0\in D$ of $\calH$ is finite-dimensional. Let
$P_{z_0}$ be the corresponding Riesz projection. See chapter 6 in
\cite{HislopSigal}, chapter III-6.4 in \cite{Kato}, or chapter XII.2
in \cite{ReSi4} for a definition and a discussion of the general
properties of Riesz projections. By the discussion on page 178 in
\cite{Kato} one knows that $\ran(P_{z_0})$ is a reducing subspace
for $\calH$ and one knows, see III-6.5 in \cite{Kato}, that
$\calH-z_0$ restricted to $\ran(P_{z_0})$ is quasi-nilpotent, that
is, its spectral radius is zero. Again, from formula III-6.32 in
\cite{Kato} one knows that $P_{z_0}$ is the residue of
$(\calH-z)^{-1}$ at $z=z_0$.

By the analytic Fredholm theorem, the residues of
$(1+V(\calH_0-z)^{-1})^{-1}$ at $z=z_0$ are finite rank and, using
\eqref{eq:resolventformula}, we then know in addition that $P_{z_0}$
is a finite rank operator.  In particular, $\calH-z_0$ restricted to
$\ran(P_{z_0})$ is nilpotent, since every finite rank
quasi-nilpotent operator is nilpotent, see problem I-5.6 on page 38
in \cite{Kato}. That is, there is an $m\in \N$ such that
$\ker(\calH-z_0)^m=\ran(P_{z_0})$.

The symmetry of the spectrum around the real and imaginary axis is
well-known. It follows from the fact that $\calH$ is unitary
equivalent to its adjoint $\calH^*$ and to $-\calH$. Indeed, writing
$L= -\Delta +\mu +U$, that is, $\calH=\left(\begin{smallmatrix} L &
W \\ -W & -L\end{smallmatrix}\right) $, one has
 \bdm
\left(
\begin{array}{cc}
      1 & 0 \\
      0 & -1
      \end{array}
    \right)
\left( \begin{array}{cc}
      L & W \\
      -W & -L
      \end{array}
    \right)
\left( \begin{array}{cc}
      1 & 0 \\
      0 & -1
      \end{array}
    \right)
=
\left( \begin{array}{cc}
      L & -W \\
      W & -L
      \end{array}
    \right)
= \calH^*
 \edm
and
 \bdm
\left( \begin{array}{cc}
      0 & 1 \\
      1 & 0
      \end{array}
    \right)
\left( \begin{array}{cc}
      L & W \\
      -W & -L
      \end{array}
    \right)
\left( \begin{array}{cc}
      0 & 1 \\
      1 & 0
      \end{array}
    \right)
=
\left( \begin{array}{cc}
      -L & -W \\
      W & L
      \end{array}
    \right)
= - \calH .
 \edm \qed

\begin{remark} There is an alternative way to show that
$\sigma_{\ess,5}(\calH)= \sigma(\calH)$. Once one knows that
$\rho(\calH)\not=\emptyset$ one can use Remark
\ref{rem:essentialspectrum6} and the fact that the unperturbed
operator $\calH_0$ is self-adjoint with a gap in its spectrum to
argue as follows: A simple calculation, using
\eqref{eq:resolventformula},  gives
 \bdm
  (\calH-z)^{-1} - (\calH_0-z)^{-1}
  =
  - (\calH_0-z)^{-1} (1+V(\calH_0-z)^{-1})^{-1}V(\calH_0-z)^{-1},
 \edm
for $z\in\rho(\calH)\cap\rho(\calH_0)$. Since the right hand side is
a compact operator, a version of  Weyl's criterion for suitable
non-self-adjoint operators, Theorem XIII.14 in \cite{ReSi4}, shows
$\sigma(\calH)\setminus \sigma_{\text{disc}}(T)= \sigma(\calH_0))$.
In addition, this immediately gives that the spectrum of $\calH$
outside of $\sigma(\calH_0)$ has finite algebraic multiplicity since
it is the discrete spectrum.
\end{remark}

\section{Exponential decay of eigenfunctions} \label{sec:Ltwo-decay}
We show in this section that every eigenfunction of $\calH$
corresponding to an eigenvalue $E$ with $-\mu<\re E < \mu$ decays
exponentially.

Let $\varphi = \left(\begin{smallmatrix} \varphi_1 \\ \varphi_2 \end{smallmatrix}\right)$
be an eigenfunction with eigenvalue $E$, i.e., $\mathcal{H} \varphi = E \varphi$, or,
    \begin{align*}
    L \varphi_1 + W \varphi_2 &= E \varphi _1\\
    - W \varphi_1 - L \varphi _2 &= E \varphi _2.
    \end{align*}
This can be rewritten as
    $$
    \left(  \begin{array}{cc}
     L-E & W \\
     W & L+E
      \end{array}
    \right)
    \left( \begin{array}{c}
    \varphi _1\\
    \varphi _2
    \end{array}
    \right)=0.
    $$
Thus we are led to study the zero energy eigenfunctions of the
energy dependent operator
    $$
    \hatt{H}_E:=\left(  \begin{array}{cc}
     L-E & W \\
      W & L+E
      \end{array}
    \right).
    $$
Recalling $L= -\Delta+\mu+U$, we can write $\hatt{H}_E=\hatt{H}_{0,E}+\hatt{V}$ with
    $$
    \hatt{H}_{0,E}:=\left(  \begin{array}{cc}
     -\Delta + \mu -E & 0 \\
      0 & -\Delta + \mu +E
      \end{array}
    \right)
    \quad \text{and} \quad
    \hatt{V}:= \left( \begin{array}{cc}
    U & W \\
    W & U
    \end{array}
    \right).
    $$
Note that the eigenvalue $E$ need not be real, since $\calH$ is not a self-adjoint
operator. This corresponds to the fact that, for complex $E$, the energy dependent
operator $\hatt{H}_E$ will also not be self-adjoint. In this case, we have
$\hatt{H}_E = \re \hatt{H}_E + i \im \hatt{H}_E$, where
    $$\re \hatt{H}_E= \left( \begin{array}{cc}
        -\Delta + \mu -\re E + U & W \\
      W & -\Delta + \mu +\re E  + U
      \end{array}
    \right)
    $$
and $$ \im \hatt{H}_E = \left( \begin{array}{cc}
        -\im E & 0 \\
      0 & \im E
      \end{array}
    \right).
    $$
To prove exponential decay of $\varphi _1$ and $\varphi _2$, we
apply a modification of the Agmon method from the theory of
Schr\"{o}dinger operators, see, for example, \cite{HS}, to the
operator $\hatt{H}_E$. We need the following three preparatory lemmas.

\begin{lemma}\label{Lemma1} Let $B_R^c= \{x\in\R^d : |x| \ge R \}$. Then
    \begin{equation}\label{eq:Sigma}
    \Sigma := \lim _{R \to \infty} \inf \Big\{ \frac{ \langle \varphi, \re(\hatt{H}_E) \varphi \rangle}{\| \varphi\|^2} :
     \varphi \in H^2(\R^d, \C^2),\ \supp (\varphi) \subset B_R^c \Big\} \geq \mu_E,
    \end{equation}
where we put $\mu_E := \mu-|\re E|$.
\end{lemma}

\begin{proof}
Since $-\Delta \geq 0$, we obtain $\re\hatt{H}_{0,E} \geq \mu_E$.
Indeed, for any $\varphi \in Dom(\hatt{H}_0)=H^2(\R^d,\C^2)$,
    \begin{align}\label{hatH_0}
    \langle \varphi , \re\hatt{H}_{0,E} \varphi \rangle _{L^2(\R^d, \C^2)}
    &=
      \langle \varphi_1, (-\Delta +\mu -\re E)\varphi_1 \rangle_{L^2(\R^d)} \notag \\
    & \phantom{=} +\langle \varphi_2, (-\Delta +\mu +\re E)\varphi_2\rangle_{L^2(\R^d)} \notag \\
    & \geq (\mu -\re E)\|\varphi_1\|_{L^2}^2+(\mu +\re E)\|\varphi_2\|_{L^2}^2  \notag\\
    & \geq \mu_E \| \varphi \|_{L^2(\R^d, C^2)}^2.
    \end{align}
To estimate $\langle \varphi, \hatt{V} \varphi \rangle$, note that the matrix
$\hatt{V} = \left( \begin{smallmatrix} U & W \\ W &U \end{smallmatrix} \right)$
has eigenvalues $U\pm |W|$. Thus
\bdm
\langle \varphi, \hatt{V}\varphi \rangle_{L^2(\R^d, \C^2)}
\ge
\int_{\R^d} (U(x) -|W(x)|)(|\varphi_1(x)|^2 +|\varphi_2(x)|^2)\, dx .
\edm

By assumption A, the two functions $U$ and $W$ tend to zero at
infinity, so for any $\veps >0$, there exists $R_{\epsilon} > 0$
such that $U(x)\ge - \veps/2$ and $|W(x)|< \veps/2$ whenever
$|x|>R_{\epsilon}$. Using this and the above lower bound, one
immediately gets for any $\varphi$ with $\supp (\varphi) \subset
B_{R_{\epsilon}}^c$,
 \beq\label{eq:V}
    \langle \varphi ,\hatt{V} \varphi \rangle  \ge -\epsilon \| \varphi \|_{L^2(\R^d,\C^2)}^2.
 \eeq
Combining \eqref{hatH_0} and \eqref{eq:V}, we get
    \beq\label{eq:lower-bound}
    \re \langle \varphi, \hatt{H}_E \varphi \rangle =
    \langle \varphi, \re \hatt{H}_{0,E} \varphi \rangle +
    \langle \varphi , \hatt{V}  \varphi \rangle \geq (\mu_E -\epsilon) \| \varphi \|^2,
    \eeq
for any $\varphi \in Dom(\hatt{H})$ with $\supp (\varphi) \subset
B_{R_{\epsilon}}^c$. Since the infimum in the right-hand side of
\eqref{eq:Sigma} is increasing in $R$, \eqref{eq:lower-bound} gives
    $$
    \Sigma  \geq \mu_E-\epsilon.
    $$
Since $\epsilon >0$ is arbitrary, we conclude \eqref{eq:Sigma}.
\end{proof}

For the next lemma, we need a cut-off function $j_R$. Let $0\le j\le
1$ with $j\in \calC^\infty(\R_+)$ and $j(t)=1$ for $0\le t\le 1$ and
$j=0$ for $t\ge 2$ and put $j_R(t) = j(t/R)$. Moreover, let $\langle
x\rangle = \sqrt{1+|x|^2}$.

\begin{lemma}\label{Lemma2}
Let $-\mu<\re E<\mu$. Then for any positive $\delta <\mu_E/2$, there
exists $R=R(\delta)>0$ such that with the cut-off function $j=j_R$
and uniformly in $\veps>0$
$$
\langle j \varphi, \bigl( \re \hatt{H}_E-|\nabla f_{\epsilon}|^2
\bigr) j \varphi \rangle \geq \delta \langle j \varphi, j \varphi
\rangle,\ \varphi \in H^2(\R^d, \C^2),
$$
where $f_{\veps}(x)=\dfrac{\beta \langle x \rangle}{1+\veps
\langle x \rangle}$ with $\beta = \sqrt{\mu_E-2\delta}$.
\end{lemma}

\begin{proof} By assumption,
$\mu_E= \mu- |\re E|>0$. Pick any $0<\delta <\mu_E/2$. By
Lemma~\ref{Lemma1}, there exists $R_{\delta}
> 0$ such that for all $\varphi \in Dom(\hatt{H}_E)$ with $\supp
(\varphi)\subset B_{R_{\delta}}^c$,
    $$ \langle \varphi, \re\hatt{H}_E \varphi \rangle \geq (\mu_E-\delta)
    \langle \varphi , \varphi \rangle.$$
Thus, with the cut-off function $j_R= J_R(\delta)$, we obtain, for any
$\varphi \in H^2(\R^d, \C^2)$,
    $$ \langle j_R \varphi, \re \hatt{H}_E j_R \varphi \rangle \geq (\mu_E-\delta)
    \langle j_R \varphi , j_R \varphi \rangle.$$
Since $| \nabla f_{\epsilon} | \leq \beta $, we get
    $$ \langle j_R \varphi, \bigl( \re\hatt{H}_E-|\nabla f_{\epsilon}|^2 \bigr) j_R \varphi\rangle
    \geq (\mu_E-\delta -\beta ^2) \langle j_R \varphi, j_R \varphi \rangle
    = \delta \langle j_R \varphi, j_R \varphi \rangle .
    $$
\end{proof}

\begin{lemma}\label{Lemma3}
If, in addition to the hypothesis of Lemma~\ref{Lemma2}, $\hatt{H}_E
\varphi =0$, then
    \begin{equation}\label{E:Lemma3-0}
    \|j_R e^{f_{\epsilon}} \varphi \| \leq \delta ^{-1} \|e^{f_{\epsilon}} [\hatt{H}_0, j_R] \varphi
    \|
    \end{equation}
where $[\hatt{H}_0, j_R]= \hatt{H}_0 j_R - j_R \hatt{H}_0$ and
$\hatt{H}_0=\left(\begin{smallmatrix} -\Delta + \mu & 0 \\ 0 &
-\Delta + \mu\end{smallmatrix}\right)$.
\end{lemma}
\begin{proof} Let $\calC^\infty_b(\R^d)$ be the set of bounded, infinitely
often differentiable functions.
Note $e^{\pm g} \calD(\hatt{H}_E) = \calD(\hatt{H}_E)$ and, since
$e^g \hatt{H}_Ee^{-g}= e^g \re \hatt{H}_Ee^{-g} + i \im \hatt{H}_E$, also
    \begin{equation}\label{E:Lemma3-1}
    \re \langle \psi, e^g \hatt{H}_E e^{-g} \psi \rangle =
    \langle \psi, \bigl( \re\hatt{H}_E-|\nabla g |^2 \bigr) \psi \rangle
    \end{equation}
for any $\psi \in Dom(\hatt{H}_E)$ and any real valued function
$g\in\calC^\infty_b(\R^d)$, see Appendix \ref{app:A}.

Let $\hatt{H}_E \varphi =0$. Since $f_\vare \in\calC^\infty_b(\R^d)$,
the product $e^{f_{\veps}}\varphi$ is in the domain of $\hatt{H}_E$.
So we can apply Lemma~\ref{Lemma2} with $\varphi$ replaced by
$e^{f_{\epsilon}} \varphi$. Using \eqref{E:Lemma3-1}, we obtain
    \begin{align}\label{E:Lemma3-2}
    \delta \|j_R e^{f_{\epsilon}} \varphi \|^2 & \leq
    \langle j_R e^{f_{\epsilon}} \varphi,
    \bigl( \re\hatt{H}_E-|\nabla f_{\epsilon} |^2 \bigr) j_R e^{f_{\epsilon}} \varphi
    \rangle \notag \\
    &= \re  \langle j_R e^{f_{\epsilon}} \varphi,
       e^{f_{\epsilon}} \hatt{H}_E e^{-f_{\epsilon}} j_R e^{f_{\epsilon}} \varphi
    \rangle \notag \\
    &= \re  \langle j_R e^{f_{\epsilon}} \varphi, e^{f_{\epsilon}} \hatt{H}_E j_R \varphi
    \rangle.
    \end{align}
As $\hatt{H}_E \varphi =0$, the right hand side of
\eqref{E:Lemma3-2} is equal to $ \re \langle j e^{f_{\epsilon}}
\varphi, e^{f_{\epsilon}} [ \hatt{H}_E, j ] \varphi \rangle$. Then,
by the Cauchy-Schwarz inequality and
$[\hatt{H}_E,j]=[\hatt{H}_0,j]$, we conclude \eqref{E:Lemma3-0}.
\end{proof}

The following corollary finishes the proof of Theorem \ref{thm:Ltwo-exp-decay}
for eigenfunctions.

\begin{corollary}[$=$Theorem \ref{thm:Ltwo-exp-decay} for eigenfunctions]\label{cor:1}
Let $-\mu<\re E<\mu$ and $\varphi$ an eigenfunction of zero energy
for $\hatt{H}_E$, i.e., $\mathcal{H} \varphi =E
\varphi$. Then $\varphi$ decays exponentially. More precisely, for
all positive $\delta < \frac{1}{2}(\mu -|\re E|)$, \bdm
e^{(\sqrt{\mu-|\re E|-2\delta})|x|} \varphi(x)\in L^2(\R^d,\C^2).
\edm
\end{corollary}
\begin{proof}
Simply note that $[\hatt{H}_{0},j]$ is a first order differential
operator concentrated on the annulus $R\le |x|\le 2R$. Indeed,
 \bdm
 \begin{split}
 [\hatt{H}_{0}, j_R] &=
     \left( \begin{array}{cc}
     [-\Delta+\mu, j_R] & 0 \\
      0 & [-\Delta+\mu, j_R]
      \end{array} \right)
      =
      \left( \begin{array}{cc}
     [-\Delta, j_R] & 0 \\
      0 & [-\Delta, j_R]
      \end{array} \right)\\
   & =
    \left(\begin{array}{cc}
     (-\Delta j_R)- \nabla j_R \cdot \nabla & 0 \\
      0 & (-\Delta j_R)- \nabla j_R \cdot \nabla
      \end{array} \right)
 \end{split}
 \edm
and $j_R$ is constant outside the annulus $R\le |x|\le 2R$. Thus
$e^{f_{\epsilon}}[-\Delta,j]$ is a bounded operator in $H^2(\R^d)$
with a uniform bound in $\epsilon$. Hence also \bdm
\limsup_{\vare\to 0} \| e^{f_{\epsilon}} [\hatt{H}_{0}, j] \varphi
\| <\infty \edm for any $\varphi\in H^2(\R^d,\C^2)$. Since
$f_\veps\uparrow f$ as $\veps\to 0$ we can use dominated convergence
and \eqref{E:Lemma3-0} to conclude for any eigenfunction of $\calH$
with energy $E$ \bdm \|e^{\sqrt{\mu_E- 2\delta}\langle x\rangle} j
\varphi \| = \lim_{\vare\to 0 } \| e^{f_\vare} j\varphi\ \| <\infty.
\edm for any $0<\delta < \mu_E /2$, where $\mu_E = \mu- |\re E|>0$,
by assumption. Since $j=1$ outside a compact set, $e^{\sqrt{\mu_E-
2\delta}\langle x\rangle} \varphi$ is square integrable on all of
$\R^d$.
\end{proof}

\section{Exponential decay of generalized eigenfunctions} \label{sec:Ltwo-decay-generalized}
The method in the previous section can be used to show that
all generalized eigenfunctions decay exponentially. We need a
little extension of Lemma \ref{Lemma3}. But first some
more notation: For $\varphi \in L^2(\R^d,\C^2)$ let
$\wti{\varphi} = \left(\begin{smallmatrix} \varphi_1\\-\varphi_2\end{smallmatrix}\right)$.
With this, we have the following

\begin{lemma}\label{lem:gen-decay-estimate} Let $-\mu <\re E<\mu$.
Assume that for some $k\in\N$,  $\psi_{l-1}\in
Dom(\hatt{H}_E)^{k-(l-1)}$ for $l= 1,\ldots,k$ with
$\tilde{\psi_l} = \hatt{H}_E\psi_{l-1}$. Then for all positive
$\delta<\mu_E/2$ we have \beq\label{eq:gen-decay-estimate} \|j
e^{f_\vare}\psi_0\| \le \sum_{l=0}^{k-1} \delta^{-(l+1)}\|
e^{f_\vare}[\hatt{H}_0,j]\psi_l \| + \delta^{-k}\|
e^{f_\vare}j\psi_k\| \eeq
\end{lemma}
\begin{proof}
It is enough to show that
\beq\label{eq:gen-decay-1}
    \|j e^{f_\vare}\psi_{l-1}\|\le
    \delta^{-1} \big(
    \| e^{f_\vare}[\hatt{H}_0,j]\psi_{l-1} \| + \| e^{f_\vare}j\psi_l \|
    \big)
\eeq
for $l=1,\ldots,k$. Then \eqref{eq:gen-decay-estimate} follows from iterating this bound.

Using Lemma~\ref{Lemma2} and the assumption $\hatt{H}_E\psi_{l-1}=\wti{\psi}_l$ we see
    \begin{align*}
   \delta \|j e^{f_{\epsilon}} \psi_{l-1} \|^2 &
   \leq \langle j e^{f_{\epsilon}} \psi_{l-1}, \bigl( \re \hatt{H}_E-|\nabla f_{\epsilon} |^2 \bigr) j e^{f_{\epsilon}} \psi_{l-1}
    \rangle\\
   &=\re  \langle j e^{f_{\epsilon}} \psi_{l-1}, e^{f_{\epsilon}} \hatt{H}_E e^{-f_{\epsilon}} j e^{f_{\epsilon}} \psi_{l-1}
    \rangle \notag \\
    &= \re  \langle j e^{f_{\epsilon}} \psi_{l-1}, e^{f_{\epsilon}} \hatt{H}_E j \psi_{l-1}
    \rangle \\
    &= \re  \langle j e^{f_{\epsilon}} \psi_{l-1}, e^{f_{\epsilon}}
    [\hatt{H}_E, j] \psi_{l-1} +e^{f_{\epsilon}}j \wti{\psi}_l
    \rangle\\
    &=\re  \langle j e^{f_{\epsilon}} \psi_{l-1}, e^{f_{\epsilon}}
    [\hatt{H}_0, j] \psi_{l-1} \rangle +\re \langle j e^{f_{\epsilon}} \psi_{l-1},
    je^{f_{\epsilon}} \wti{\psi}_{l} \rangle\\
    & \leq \| j e^{f_{\epsilon}} \psi_{l-1} \| \Bigl\{ \bigl\| e^{f_{\epsilon}}
    [\hatt{H}_0, j] \psi_{l-1} \bigr\| + \| j e^{f_{\epsilon}} \wti{\psi}_l \|
    \Bigr\} ,
   \end{align*}
which gives  \eqref{eq:gen-decay-1}, since
$\|  j e^{f_{\epsilon}} \wti{\psi}_l \| = \|  j e^{f_{\epsilon}} \psi_l \| $.
\end{proof}

\begin{corollary}[$=$Theorem \ref{thm:Ltwo-exp-decay} for generalized eigenfunctions]\label{cor:2}
Let $E\in \C$ with  $\mu<\re E<\mu$ and $\varphi$ be a generalized
eigenfunction of $\calH$ with eigenvalue $E$. Then, for all positive
$\delta<\frac{1}{2}(\mu-|\re E|)$, \bdm e^{(\sqrt{\mu-|\re
E|-2\delta})|x|}\varphi \in L^2(\R^d,\C^2). \edm
\end{corollary}

\begin{proof} Using Theorem \ref{thm:spectrum}, see also
Remark \ref{rem:gen-eigenspace}, we know that the generalized
eigenspace corresponding to $E$ is finite dimensional. Thus there is
a $k\in \N$ such that $\ker(\calH-E)^{m+1} = \ker(\calH-E)^m$ for
all $m\ge k$. So fix this $k$ and assume that $(\calH-E)^k\varphi =
0$. Put $\psi_l = (\calH-E)^l\varphi$ and $\psi_0= \varphi$. Then
$\psi_l= (\calH-E)\psi_{l-1}$. Or, in terms of the operator
$\hatt{H}_E$, \bdm \wti{\psi}_l = \hatt{H}_E\psi_{l-1}. \edm Note
that in this case $\psi_k=0$. So for all $0<\delta<\mu_E/2$ and
large enough $R$ Lemma \ref{lem:gen-decay-estimate} gives for all
$\vare>0$ \bdm \|j e^{f_\vare}\psi_0\| \le \sum_{l=0}^{k-1}
\delta^{-(l+1)}\| e^{f_\vare}[\hatt{H}_0,j]\psi_l \| . \edm Letting
$\vare\to 0$, as in the proof of Corollary \ref{cor:1}, finishes the
proof.
\end{proof}

\begin{appendix}
\section{Proof of equation \eqref{E:Lemma3-1}} \label{app:A}

Here we prove equation \eqref{E:Lemma3-1}, that is,
    $\re \langle \psi, e^g \hatt{H}_E e^{-g} \psi \rangle =
    \langle \psi, \bigl( \re\hatt{H}_E-|\nabla g |^2 \bigr) \psi
    \rangle$
    for any $\psi \in Dom(\hatt{H}_E)$ and any real valued function
$g\in\calC^\infty_b(\R^d)$.

\begin{proof}
Since $\nabla (e^{-g} \psi)=e^{-g} \bigl( \nabla \psi -( \nabla g
) \psi \bigr)$, we have $e^g \nabla e^{-g}= \nabla - \nabla g$. Thus,
    \bdm
    \begin{split}
    e^g (- \Delta)e^{-g}& =-(e^g \nabla e^{-g})^2
    =-(\nabla -\nabla g)^2
    = -\Delta + \nabla \cdot \nabla g +\nabla g \cdot \nabla -(\nabla g)^2\\
    &=
     -\Delta -| \nabla g|^2+iB
    \end{split}
    \edm
where the operator $B= -i(\nabla \cdot \nabla g + \nabla g\cdot \nabla)$
is self-adjoint. Therefore,
    \begin{align*}
    e^g \hatt{H}_E e^{-g} &= \left( \begin{array}{cc}
                                e^g(-\Delta)e^{-g}+\mu -E & 0 \\
                                0 & e^g(-\Delta)e^{-g}+\mu +E
                                \end{array}
                        \right)+e^g \hatt V e^{-g}\\
                        &=\left( \begin{array}{cc}
                                -\Delta -| \nabla g|^2+iB +\mu -E & 0 \\
                                0 & -\Delta -| \nabla g|^2+iB +\mu -E
                                \end{array}
                        \right)+\hatt V \\
                        &=\hatt{H}_E - |\nabla g|^2+iB.
    \end{align*}
Taking the real part, one arrives at \eqref{E:Lemma3-1}.
\end{proof}

\section{On the equality of certain essential spectra} \label{app:B}
Let us now prove \eqref{eq:equivalence}, that is,
$\sigma(T)\setminus\sigma_{\disc}(T)= \sigma_{\ess,5}(T)$ for any
closed operator $T$ in a Banach space $X$. Here $\sigma(T)$ is the
complement of the resolvent set $\rho(T)$ and the discrete spectrum
$\sigma_{\disc}(T)$ is the set of all isolated points in $\sigma(T)$
with finite \emph{algebraic} multiplicity. Recall that in this case,
the nullity and deficiency are given by
 \bdm
 \nul(T)= \dim\ker(T)
 \edm
and
 \bdm
 \de(T) = \dim \big(X/\ran(T)\big) .
 \edm
From the definition of resolvent set, the set of all $z\in\C$ for
which $T-z$ is a bijection (and hence boundedly invertible, by the
inverse theorem), we have
 \bdm
 \rho(T)= \{ z\in\C\vert\, T-z \text{ is Fredholm with } \text{nul}(T-z)= \text{def}(T-z)=0
 \} .
 \edm
Recall that $\sigma_{\ess,5}(T)= \C\setminus \Delta_5(T)$ with
 \bdm
 \begin{split}
 \Delta_5(T)= \{ z\in\C\vert\, & T-z  \text{ is Fredholm with } \ind(T-z)= 0 \\
                                & \text{ and a deleted neighborhood of }z \text{ is in
 }\rho(T)\}.
 \end{split}
 \edm
A straightforward rewriting of this condition shows that
 \bdm
 \begin{split}
 \Delta_5(T)= \rho(T)\cup
                \{ \lambda \vert\, & \lambda \text{ is an isolated point in }\sigma(T) \text{ such that } \\
                                    & T-\lambda \text{ is Fredholm with } \ind(T-\lambda)= 0 \}.
 \end{split}
 \edm
Thus to show \eqref{eq:equivalence} it is enough to prove
\begin{lemma} \label{lem:equivalence} Let $T$ be a closed operator on some Banach space $X$.
Then
 \bdm
 \begin{split}
 \sigma_{\disc}(T)= \{ \lambda \vert\, & \lambda \text{ is an isolated point in }\sigma(T) \text{ such that } \\
                                    & T-\lambda \text{ is Fredholm with } \ind(T-\lambda)= 0 \}.
 \end{split}
 \edm
\end{lemma}

\begin{proof}
Let $\lambda$ be an isolated point in $\sigma(T)$  such that
$T-\lambda$ is Fredholm with index zero. Since $\lambda \not\in
\rho(T)$ and $\ind(T-z)= 0$, we must have $0 < \nul(T-\lambda)<
\infty$, which implies $\lambda$ is an eigenvalue of $T$ with finite
\emph{geometric} multiplicity. Since $\ran(T-\lambda)$ is closed,
Theorem IV-5.10 in conjunction with Theorem IV-5.28 in \cite{Kato}
shows that the algebraic multiplicity of $\lambda$ must also be
finite. Hence $ \lambda\in \sigma_{\text{disc}}(T)$.

Conversely, let $\lambda \in \sigma_{\disc}(T)$, i.e., $\lambda$ be
an isolated point in $\sigma(T)$ with finite \emph{algebraic}
multiplicity. We need to show that $T-\lambda$ is Fredholm with
index zero. By Theorem III-6.17, together with Section III-6.5 in
\cite{Kato}, there is a decomposition of $X=M'\oplus M''$ such that
$M'$ and $M''$ are reducing subspaces for $T$ and $M'\cap M''={0}$.
In fact, if $P_\lambda$ is the Riesz projection corresponding to
$\lambda$, then $M'= \ran(P_\lambda)$ and $M''= \ran(1-P_\lambda)$.
Furthermore, $T-\lambda$ restricted to $M'$ is bounded and
quasi-nilpotent and $T-\lambda$ restricted to $M''$ is bijective.
Note that $\lambda$ having finite algebraic multiplicity is
equivalent to $M'$ being finite-dimensional. In particular,
$\ran(T-\lambda)= \ran((T-\lambda)\vert_{M'})\oplus M''$ is closed.
One has
 \bdm
 \ker(T-\lambda)= \ker((T-\lambda)\vert_{M'}) \oplus
 \ker((T-\lambda)\vert_{M''})
 \edm
and
 \bdm
 X/(T-\lambda)= M'/(T-\lambda)\vert_{M'} \oplus
 M''/(T-\lambda)\vert_{M''} .
 \edm
Hence,
 \bdm
 \nul(T-\lambda)= \nul((T-\lambda)\vert_{M'})+
\nul((T-\lambda)\vert_{M''})
 \edm
and
 \bdm
 \de(T-\lambda)= \de((T-\lambda)\vert_{M'})+
 \de((T-\lambda)\vert_{M''}) .
 \edm
Since $(T-\lambda)\vert_{M''}$ is a bijection, we know that the
second terms above are zero, that is, $ \nul(T-\lambda)=
\nul((T-\lambda)\vert_{M'}) $ and $ \de(T-\lambda)=
\de((T-\lambda)\vert_{M'})$. Moreover, both are finite, since $M'$
is finite dimensional, and the well-known dimension formula from
finite dimensional linear algebra shows that
    \bdm
    \nul((T-\lambda)\vert_{M'}) = \de((T-\lambda)\vert_{M'}) .
    \edm
Thus $T-\lambda$ is indeed Fredholm with index zero.

\end{proof}

\vspace{0.5em}
\begin{remark} Let us also remark on the original definition of
essential spectrum by Browder, see Definition 11 on page 107 in
\cite{Browder}. Browder defines $\sigma_{\ess,B}(T)= \C\setminus
\Delta_B(T)$ with
 \bdm
 \begin{split}
 \Delta_B(T)= \{z\in C\vert\, & \ran(T-z) \text{ is closed, } z \text{ is of finite algebraic multiplicity, }\\
                            & \text{ and }z\text{ is not a limit point of } \sigma(T)
                            \} .
 \end{split}
 \edm
One can rewrite this as
 \bdm
 \begin{split}
 \Delta_B(T)= \rho(T)\cup \{z\in C\vert\, & \ran(T-z) \text{ is closed and } z \text{ is an isolated point } \\
                            & \text{ in }\sigma(T)\text{ with finite algebraic multiplicity}
                            \} .
 \end{split}
 \edm
As shown in the proof of Lemma \ref{lem:equivalence}, for any
isolated point $z\in\sigma(T)$ with finite algebraic multiplicity,
$\ran(T-z)$ is always closed. Thus, in fact,
 \bdm
 \begin{split}
 \Delta_B(T) & = \rho(T)\cup \{z\in C\vert\,  z \text{ is an isolated point in }\sigma(T)\\
            & \phantom{\rho(T)\cup \{z\in C\vert abc \}} \text{ with finite algebraic multiplicity}\} \\
            & = \rho(T) \cup \sigma_{\disc}(T) ,
 \end{split}
 \edm
where the last equality is due to Lemma \ref{lem:equivalence}. Hence
Browder's original definition indeed gives the same essential
spectrum as $\sigma_{\ess,5}(T)$.
\end{remark}

\end{appendix}

\renewcommand{\thesection}{\arabic{chapter}.\arabic{section}}
\renewcommand{\theequation}{\arabic{chapter}.\arabic{section}.\arabic{equation}}
\renewcommand{\thethm}{\arabic{chapter}.\arabic{section}.\arabic{thm}}

\def\cprime{$'$}

\end{document}